%% file: main.tex
\documentclass[sigconf]{acmart}

\usepackage{balance} % For balanced columns on the last page
\usepackage{booktabs}

\setcopyright{rightsretained}
\copyrightyear{2017}

\begin{document}
\title[EpiFi]{EpiFi: An In-Home Sensor Network Architecture\\for Epidemiological Studies}

\author{Philip Lundrigan}
\affiliation{University of Utah}
\email{philipbl@cs.utah.edu}

\author{Kyeong Min}
\affiliation{University of Utah}
\email{kyeong.min@utah.edu}

\author{Neal Patwari}
\affiliation{University of Utah}
\email{npatwari@ece.utah.edu}

\author{Sneha Kasera}
\affiliation{University of Utah}
\email{kasera@cs.utah.edu}

\author{Kerry Kelly}
\affiliation{University of Utah}
\email{kerry.kelly@utah.edu}

\author{Jimmy Moore}
\affiliation{University of Utah}
\email{jimmy@cs.utah.edu}

\author{Miriah Meyer}
\affiliation{University of Utah}
\email{miriah@cs.utah.edu}

\author{Scott C. Collingwood}
\affiliation{University of Utah}
\email{scott.collingwood@hsc.utah.edu}

\author{Flory Nkoy}
\affiliation{University of Utah}
\email{flory.nkoy@hsc.utah.edu}

\author{Bryan Stone}
\affiliation{University of Utah}
\email{bryan.stone@hsc.utah.edu}

\author{Katherine Sward}
\affiliation{University of Utah}
\email{kathy.sward@nurs.utah.edu}

\renewcommand{\shortauthors}{P. Lundrigan et al.}

\begin{abstract}
We design and build a system called EpiFi, which allows epidemiologists to easily design and deploy experiments in homes. The focus of EpiFi is reducing the barrier to entry for deploying and using an in-home sensor network. We present a novel architecture for in-home sensor networks configured using a single configuration file and provide: a fast and reliable method for device discovery when installed in the home, a new mechanism for sensors to authenticate over the air using a subject's home WiFi router, and data reliability mechanisms to minimize loss in the network through a long-term deployment. We work collaboratively with pediatric asthma researchers to design three studies and deploy EpiFi in homes.
\end{abstract}




\maketitle

\input{parts/introduction}

\input{parts/related_work}

\input{parts/architecture}

\input{parts/deployments}

\input{parts/conclusion}

\begin{acks}
    Research reported in this paper was supported by the NIBIB of the National Institutes of Health under award number U54EB021973. The content is solely the responsibility of the authors and does not necessarily represent the official views of the National Institutes of Health.
\end{acks}

\balance
\bibliographystyle{ACM-Reference-Format}
\bibliography{biblio}

\end{document}

%% file: parts/introduction.tex
\section{Introduction}\label{introduction}

The holy grail of epidemiological research is to have continuous sensing of every person's exposures and activities alongside data on their health outcomes.  The conglomeration of the environmental effects on a person over their lifetime is called their \emph{exposome} and, in interaction with their genome, plays a large part in their health \cite{wild2012exposome}. Detailed exposome data from a segment of the population, including from sensors in their homes, could provide researchers new insights about the relationships between exposure and chronic diseases, such as heart disease, cancer, diabetes, and asthma, and how we can improve our health and reduce the incidence and costs of these diseases \cite{van2012integrated}.  Low cost internet-of-things (IoT) devices and wireless and wearable sensors are enabling the ``quantified self'' for those who are technologically skilled and most interested in self-monitoring \cite{swan2012sensor}.  However, for epidemiology researchers who wish to deploy and obtain reliable data from a large population of volunteer subjects, several challenges must be addressed.  In short, the costs and data privacy requirements for human subject research studies are fundamentally different from those for individuals who wish to monitor themselves.  Today's networking tools are focused on the individual user, rather than the researcher, and new tools are needed to enable a new kind of epidemiology research.
\begin{itemize}
    \item The individual user is driven by the cost of devices; but researchers' costs are primarily comprised of the cost of study management, including a high cost any time it is required to meet at, and travel to, the subject's home.  Deployment at a subject's home must be fast and reliable, regardless of the particular configuration of the subject's home WLAN.
    \item The individual user may be willing to store their data in multiple commercial cloud servers, regardless of data privacy protections.  However, a medical research must ensure that any deployed system abides by laws regarding medical information privacy, such as the Health Insurance portability and Accountability Act (HIPAA) in the US.
    \item If data is lost for a period of time during collection, an individual may be displeased, but the data will still be useful to them.  A researcher must analyze data from dozens or hundreds of subjects and large data loss may force them to throw out the data from the subject entirely to preserve uniformity across subjects.
\end{itemize}

As a result of these differences with standard IoT deployments, we designed and developed \emph{EpiFi}, a system that gathers data from home monitoring devices with a low burden of implementation. The goal of EpiFi is to allow epidemiologists to create rich and comprehensive experiments from start to finish, and by doing so, we hope EpiFi will gain wide adaption among epidemiologists. We focus on reducing the barrier to entry for configuring, deploying, and running a sensor network. With this focus in mind, we have found five key problems: configuration, secure network connectivity, sensor discovery, data persistence, and reliability. We built a solution that addresses each of these problems. These problems have been studied to various degrees in previous work, however, we are not aware of any system that simultaneously addresses all of these problems. EpiFi is open source and built on open source components. The source code is available at (deleted for double blind review).

To solve the secure network connectivity problem, we built a novel protocol for WiFi sensors to authenticate with a subject's home WiFi router. This protocol securely sends the subject's home network name and password to allow deployed sensors to connect to the network. It protects against common threat vectors such as packet manipulation and replay attacks. It uses erasure coding to minimize the effects of wireless packet loss. We also create an approach for data persistence and reliability which is not limited by power usage constraints. This approach ensures that all data measured by sensors is transmitted to the cloud for storage without loss.

To test our design, we deploy EpiFi in five long-term deployments for three different purposes.  We describe the key challenges that should be considered in the deployment of sensor networks for purposes of epidemiology research and how EpiFi addresses these issues.  We have worked collaboratively with researchers who study pediatric asthma in the design of EpiFi and the realizations of the system now deployed in the homes of human subject study participants.  We describe protocol design decisions and evaluate our choices quantitatively.  We show results on the collection reliability of sensor data, and feedback from researchers who have deployed our systems in human subjects research.  We conclude by addressing challenges for future research.

%Each of the following advances help with this goal. EpiFi allows a researcher to encode the sensing requirements of a study into a single text file that is used to configure a set of networking and sensing devices to be deployed in a home; provides a fast and reliable method for device discovery when installed in the home; provides a secure and easy method for the person deploying a system to authenticate each device to the home WiFi router without prior knowledge of the network name and password; and data reliability to minimize loss in the network through a long-term deployment.  EpiFi is built upon Home Assistant, which provides compatibility with a wide range of IoT sensors and actuators.  EpiFi is open source and available at (deleted for double blind review).  Key contributions of the project are:
%\begin{itemize}
%    \item This paper presents a novel architecture for in-home sensor networks to be used by researchers studying residents' exposure. A key feature of the architecture is that the configuration of the deployed system requires a single configuration file.
%    \item We provide a mechanism for devices brought by researchers to the home to be authenticated over the air to the subject's home WiFi router.
%    \item We present the systems designed for three researchers involved in pediatric asthma research and present results from the three sets of deployments.
%\end{itemize}

%% file: parts/related_work.tex
\section{Related Work}\label{related-work}

A study of past proceedings of SenSys and IPSN revealed that there has been little work done on improving the ease of deployment of sensor networks in homes. Many of the existing work focuses on energy, such as increasing battery life. In our work, we are building a complete system where the focus is not on energy savings at sensors but on reliability, data persistence, and making a system that is easy to use.

A sensor programming language called WASP~\cite{programming} has been developed focusing on application experts and not programming experts. The authors show that improving the programming language of a sensor network can have considerable impact on program accuracy and reduce development time. We share the same goal in wanting to lower the barrier of entry for application experts, but we address a different barrier. EpiFi focuses more on the infrastructure, configuration, and deployment. WASP is complementary to EpiFi and EpiFi could be improved by including a domain specific language to program sensors, rather than using Python.

Work done by T. Hnat et al.~\cite{hitchhiker} shares valuable insights and lessons learned through their many experiences doing home deployments. They discuss common problems when dealing with sensors in homes, such as power loss and  wireless connectivity issues. Such insights point to the importance of EpiFi and the numerous problems that it addresses.

There are commercial systems, like those sold by Qualcomm Life~\cite{qualcomm}, that provide similar benefits as EpiFi. This system has the appeal of being backed by a large company.  The use of cellular connectivity increases the cost compared to piggybacking onto a home's broadband connection, and researchers are typically budget-limited. Further, support for devices is limited to what a company will integrate into their system. We feel that some of the best innovations will be from new sensors that are developed and believe in the importance of open source software that can be improved upon by the community.

% Ease of use with sensors. 
% WiFi authentication
% Discovery

%% file: parts/architecture.tex
\section{EpiFi Architecture}\label{architecture}

\begin{figure}[!t]
    \centering
    \includegraphics[width=\columnwidth]{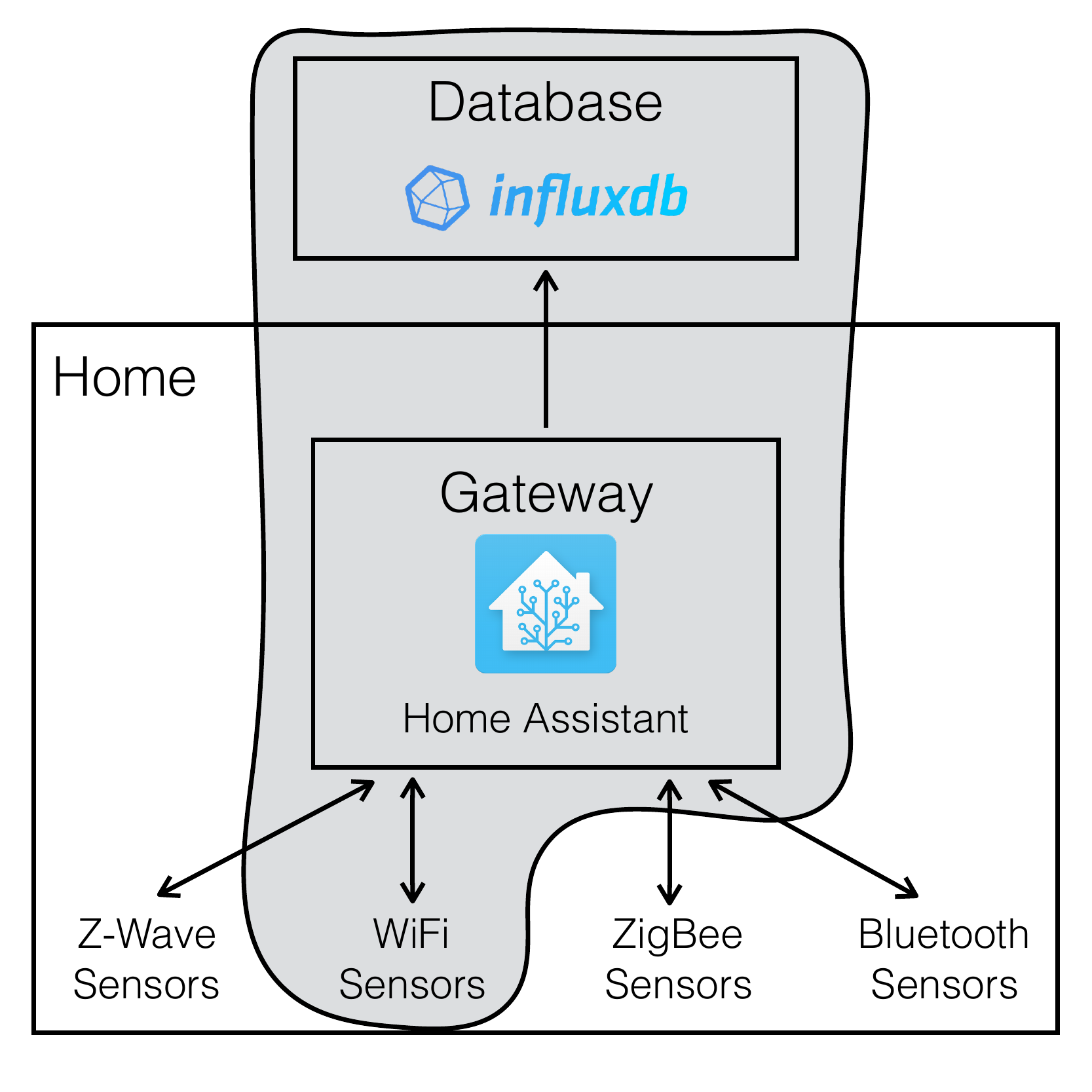}
    \caption{General architecture of EpiFi. The shaded parts are those we implement and investigate in this paper.}
    \label{fig:general_architecture}
\end{figure}

EpiFi consists of three general components, as seen in Figure~\ref{fig:general_architecture}: database, gateway, and sensors. The shaded parts are those we implement and investigate in this paper. The gateway and sensors are co-located in the home, and the database is in the cloud. The gateway acts as a central hub with all sensors and devices communicating with it. The gateway collects data from the sensors and uploads it to the database. Having all the sensors communicate with the gateway, rather than directly with the database, provides important benefits. The gateway allows the sensors to be as simple as possible because all logic, configuration, and storage for a deployment can be stored on the gateway. Also, the gateway can provide ``local reasoning'', meaning data can be acted on by the gateway such as to add privacy protections, actuation, or user feedback. Adding a gateway provides flexibility in the types of deployments that EpiFi supports, while making configuration easier. The gateway also provides the ability to interface with many different types of sensors based on protocol such as Z-Wave, ZigBee, and Bluetooth. These protocols cannot be supported unless a device in the home, like the gateway, is present.

% \begin{itemize}
%     \item The gateway can aggregate data together to save on packet overhead and round trip times. This is not a concern when dealing with most home broadband connections, but this is important when dealing with other types of wireless connections like cellular.

%     \item The gateway reduces the storage burden for the sensor when Internet access is unavailable. The gateway can store all of the data from the sensors and upload the data when a connection is available.

%     \item The gateway can provide control over the network. For example, it can pull data from sensors in a coordinated manner to help with scalability.

%     \item Only the gateway assumes the complexity of configuration. For a specific deployment, rather than having to configure each sensor individually, only one gateway needs to be configured. This dramatically reduces the time it takes to deploy such a network.

%     \item The gateway provides ``local reasoning'', meaning data can be acted on by the gateway. For example, privacy protections, actuation, or user feedback can all be implemented at the gateway rather than at particular sensors.
% \end{itemize}

To provide support for these devices, we use an open source project called Home Assistant~\cite{home-assistant}, which runs on the gateway. Home Assistant is an automation platform written in Python. It includes user contributed components that allow it to interface with devices and web services. At its core, Home Assistant is a message bus, facilitating communication among devices and functional components on a network by providing simple abstractions around home automation components such as sensors, cameras, media players, etc. For example, Home Assistant can control and collect data from popular thermostats like Nest or Ecobee or control light bulbs. As of this writing, Home Assistant supports over 600 user contributed components.

Home Assistant supports a broad range of wireless protocols such as BLE, ZigBee, Z-Wave, and WiFi. It also has a RESTful API and it supports HTTP, MQTT, raw TCP sockets, and custom components. Custom components allow users to add their own functionality without changing the core of Home Assistant. This makes it easy for new devices and sensors to integrate with Home Assistant. We develop several custom components for EpiFi.

For our remote database, we use InfluxDB~\cite{influxdb}. InfluxDB is a database designed specifically for time-series data. This fits perfectly with the sensor measurements that will be uploaded to it. To upload data to the remote database, we create a custom Home Assistant component that tags the data with a home ID before uploading. The home ID is used to uniquely identify a deployment location without compromising privacy. The database runs on a server in a campus protected environment that is HIPAA compliant. All data that is uploaded from the gateway to the server is encrypted using SSL.

Home Assistant has a large community of developers with over 470 contributors and 6,400 stars on GitHub, meaning there is abundance of documentation about the components of Home Assistant, forums and chat to get help from other users, and many blog posts and video series on how to get started. EpiFi is designed to support any device that can connect to Home Assistant, and by so doing, takes advantage of that rich knowledge base.

Although the gateway supports a broad range of wireless protocols, for the remainder of this paper, we focus on WiFi sensors because 1) WiFi hardware is readily available and inexpensive, 2) WiFi is much more widely deployed compared to the other wireless protocols~\cite{wifi-deployment}, and 3) a WiFi sensor can integrate with the rest of the home because it uses IP, making the sensor easier to debug and monitor. Any researcher can buy a commercial sensor to include in their studies, which EpiFi supports, but we feel that the most interesting work will be done with custom sensors.

For all of the reasons stated above, WiFi is a good choice for building or deploying a sensor; however, configuring a WiFi device in an established home network is a challenge. For example: How does a WiFi sensor with no direct user interface get the network name and password for a home that it is being deployed in? Once the device is connected to a wireless network, how should data be uploaded and stored to meet the privacy and reliability requirements of the researcher? EpiFi solves both of these problems.

\begin{figure}[!t]
    \centering
    \includegraphics[width=\columnwidth]{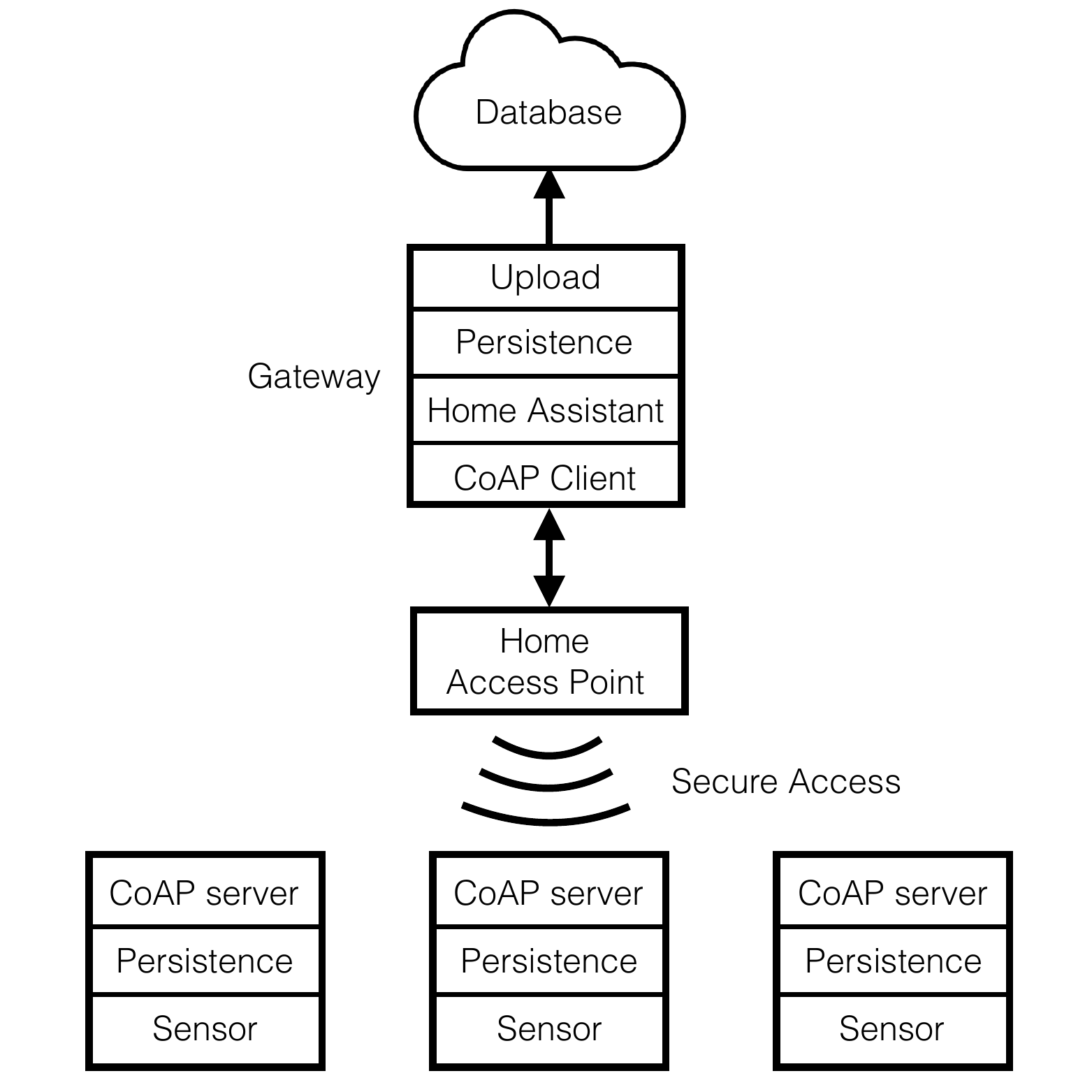}
    \caption{Architecture using WiFi sensors}
    \label{fig:wifi_architecture}
\end{figure}

Figure~\ref{fig:wifi_architecture} shows a detailed view of how WiFi sensors connect to the gateway. When connecting with WiFi sensors, we use the home access point to communicate rather than making the gateway an access point. Making the gateway an access point would provide more control over the network and make connecting WiFi sensors easier, because the network name and password can be set up before deployment; however, there are many drawbacks to using the gateway as an access point, as we discovered while testing deployments, which made this option infeasible. These experiences are shared in more detail in Section~\ref{large-deployment}, but are also summarized here. First, the commodity hardware available is not designed to be used as an access point. Some drivers support this option, but using it proved to be unreliable. A USB connected WiFi dongle is unable to take advantage of many of the features that come standard on today's access points, such as beamforming and MIMO. We also found that homes we deployed in had WiFi networks customized to the needs of the home. For example, a larger house might have multiple access points to provide complete coverage. To ensure complete coverage of the deployed sensors, we would have to essentially duplicate their network by placing multiple gateways throughput the house. This increases cost and complexity of deployment. For these reasons, we decided to utilize the home's wireless network.

To overcome the challenge of connecting WiFi sensors to a home's wireless network, we create a novel way of securely sharing the network credentials with any unassociated wireless sensors (Section~\ref{network-connectivity}) and providing automatic sensor discovery (Section~\ref{discovery}).

The last architectural decision we made is what protocol to use between sensors and gateway. There are two commonly used protocols for IoT devices: MQ Telemetry Transport (MQTT)~\cite{mqtt} and Constrained Application Protocol (CoAP)~\cite{coap}. Both of these protocols are designed as a lightweight communication protocol for power and resource constrained devices. HTTP (Hypertext Transfer Protocol), which is ubiquitously used on the Internet for communication is not well suited for these types of devices~\cite{not-http}. There have been many comparisons between MQTT and CoAP under various circumstances~\cite{mqtt_vs_coap_1}\cite{mqtt_vs_coap_2}. Based on these results, CoAP and MQTT both perform well and which one performs the best depends on the specific topology and network conditions.

We selected CoAP as the protocol between sensors and gateway because it is architecturally a better fit. The EpiFi system has a one-to-many relationship between the gateway and sensors. MQTT is a publisher-subscriber protocol that fits best with many-to-many relationships and requires a broker to receive and forward published messages. CoAP, on the other hand, is a client-server protocol with similar semantics to HTTP. This fits better with our gateway-sensor architecture.

With the basics of the architecture addressed, we decompose EpiFi into five core components.

\begin{itemize}
    \item \textbf{Configuration}: This component takes care of configuring sensors and a gateway. We focus on reducing configuration to the bare minimum and placing it in one location.

    \item \textbf{Network Connectivity}: This component deals with connecting a WiFi sensor to the home's wireless network. This is an important aspect of a deployment that is often overlooked.

    \item \textbf{Sensor and Service Discovery}: Once sensors are deployed, the gateway must be able to discover what sensors are available and what data they offer.

    \item \textbf{Data Persistence}: As said perfectly in \cite{hitchhiker}, ``homes are hazardous environments'' for sensors because random power outages and Internet disconnections are very common. Data must be aggressively stored persistently for no data to be lost.

    \item \textbf{Reliability}: Even when data is stored persistently, care must be taken to make sure no data is lost in transmission.
\end{itemize}

\subsection{Configuration}

EpiFi hides the complexity in setting up sensors and gateways controlled by Linux devices, which can be quite complicated for the non-expert. The first step in creating a deployment is setting up the sensors and gateway. Without a good understanding of how Linux works, setting up devices like these can be an insurmountable task. Small tasks like changing the hostname requires updating files in two locations (/etc/hosts and /etc/hostname). If you want to set a program to run when the device first boots up, you must figure out what ``init'' system the operating system uses (systemd, upstart, /etc/rc.local) and then learn the particularities of it. This component of EpiFi simplifies configuration.

Physical sensors in EpiFi connect to Beaglebone Black (BBB) hardware, and the hardware that a gateway runs on is a Raspberry Pi 3. We selected this hardware because it is small, cheap, has I/O pins for sensors, and is easy to develop and test on because they run Linux.

We have minimized the work to configure sensors and the gateway in three ways through EpiFi. First, we have created two custom disk images, one for sensors and one for gateways. These images come pre-configured and pre-installed with all the necessary software. Second, all configuration for a sensor or a deployment is located in one configuration file. This take the cognitive burden off of the researcher to know where each configuration file is located and how it needs to be changed. We built a script that parses the configuration file and makes the necessary changes to the system. Third, we built a generalized software component that runs on the BBB and supports user-provided sensor plug-ins. We use Python 3 as the language for these sensor plugins, in part because it is relatively easy to use and popular~\cite{python}, and there are many modules that already exist for popular sensors~\cite{python-sensors}.

We describe next the specifics of the internal configuration of sensors and gateway in EpiFi.

\subsubsection{Sensor Configuration}

Physical sensors must be selected and connected to the BBB. For example, a temperature and humidity sensor can be connected to the GPIO pins of the BBB. Second, a Python script must be written to interface and get data from the sensor. This process is beyond the scope of this paper. We present two such sensors in the Deployment section.

To integrate the sensors and code into EpiFi, the custom image for sensors must be downloaded and installed onto the SD card for the BBB. On the image there are is a configuration file called device-init.yaml. YAML~\cite{yaml} is a file format that is designed to be easily read and written by humans. All configuration for a device are stored within EpiFi in this file. For example, the hostname and password can be specified in this configuration file. When the sensor boots up, a script looks through the configuration file and changes the hostname and password to whatever is specified in device-init.yaml. The disk image also contains a folder called sensors. This is where all of the Python scripts go that interface with the specific sensors.

In order for EpiFi to support a sensor, it must be a subclass of the \texttt{Sensor} class, which EpiFi defines. This class provides a common interface for all sensors that EpiFi supports. There are two major methods that are part of this subclass, \texttt{read} and \texttt{write}. \texttt{read} obtains data from the physical sensor and then returns it. After \texttt{read} has been called on all sensors, \texttt{write} is called, which allows a sensor to display any data if needed. For example, an LCD screen can display sensor readings when its \texttt{write} method is called.

\begin{figure}[!t]
    \centering
    \includegraphics[width=\columnwidth]{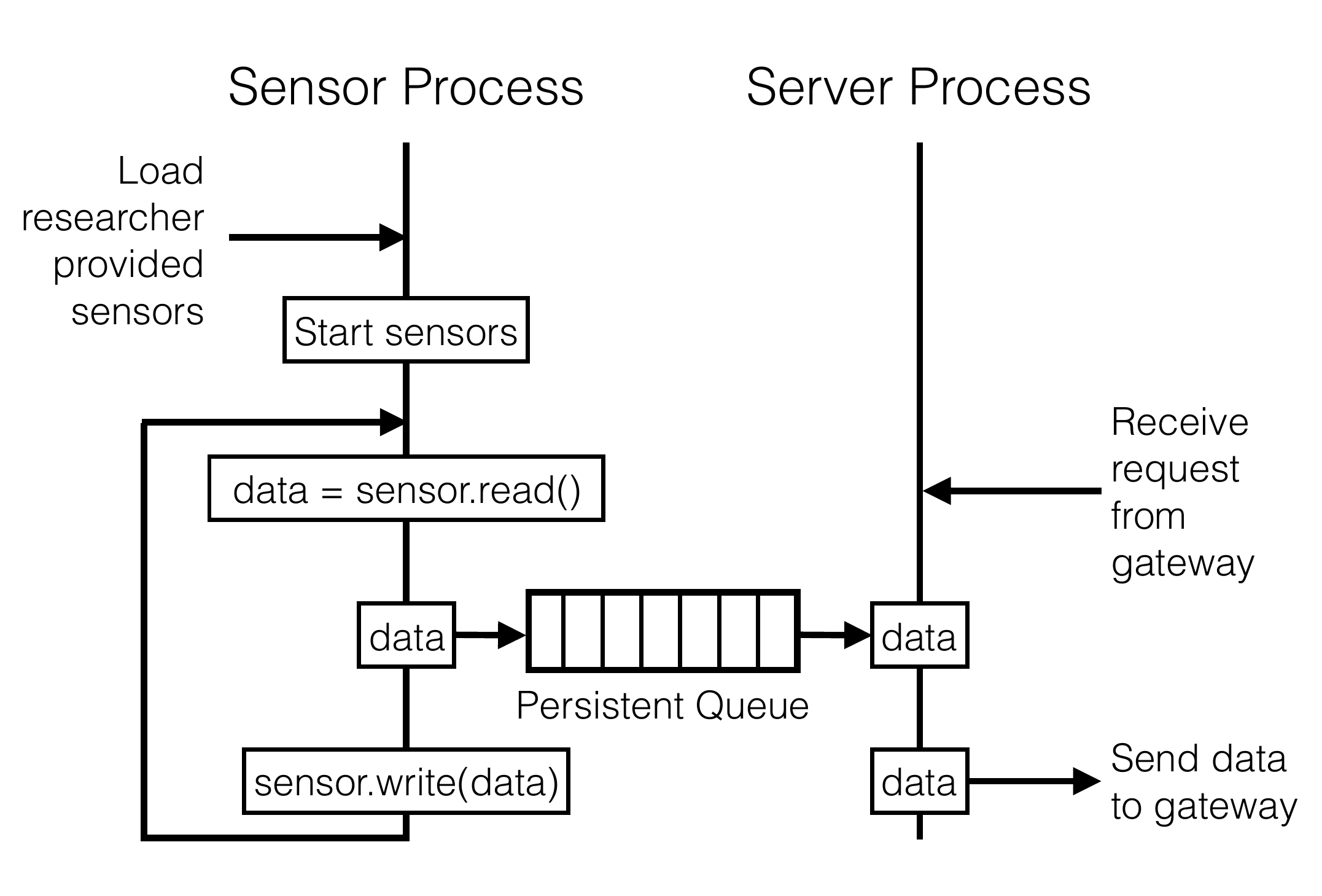}
    \caption{The two processes of the general software component. One process reads/writes data to all of the sensors and the other process services requests from the gateway.}
    \label{fig:general-software-component}
\end{figure}

The generalized software component that runs on the BBB looks for sensors in the sensor folder, starts them, and then periodically reads data from them and writes data to them. It also sets up a CoAP server that the gateway can connect to and pull data from. This is shown in Figure~\ref{fig:general-software-component}. A researcher doesn't have to worry about running the scripts or collecting data since EpiFi takes care of that for them.

\subsubsection{Gateway Configuration}

Configuring the gateway follows a similar approach, but without sensor code and more configuration options. The gateway custom image is downloaded and flashed to an SD card. In the image, there is a configuration file called device-init.yaml. This file contains all the configuration for the gateway. This is a place for researchers to configure specifics to the deployment such as the home ID where the gateway will be deployed, information about the database to which data will be uploaded (such as host, username, and password), and customizations for how often the gateway discovers and pulls data from sensors.

\subsubsection{Evaluation}

To set up and configure a gateway or sensor takes knowledge on what commands to run and tools to use for the specific operating system you are using (in EpiFi's case, Linux). Table~\ref{table:configuration} outlines some of the steps required using the default image compared to using EpiFi. In the table, a check mark means work is required and a x-mark means nothing needs to be done. Although none of these steps are difficult on their own, when put together, they can appear to be insurmountable to someone who is not familiar with computer systems. Additionally, building a gateway or sensor can be time consuming, with each taking roughly an hour to build. All of these factors get compounded when dealing with multiple deployments and multiple sensors per deployment. EpiFi's custom images reduce the amount of work and time down to that required to edit one configuration file.

\begin{table*}[!t]
\renewcommand{\arraystretch}{1.2}

\centering
\caption{Steps required to set up a sensor and gateway using a default image compared to using EpiFi. A checkmark means that it needs to be done and a xmark means it does not.}
\label{table:configuration}
\begin{tabular}{@{}lccccc@{}} \toprule

                                         & \multicolumn{2}{c}{Default Image} & & \multicolumn{2}{c}{EpiFi Image} \\
                                         & Sensor         & Gateway          & & Sensor         & Gateway        \\
                                         \cmidrule{2-3} \cmidrule{5-6}
Build sensor components                  & $\checkmark$ & $\times$     & & $\checkmark$ & $\times$     \\
Install modern version of Python (3.5+)  & $\checkmark$ & $\checkmark$ & & $\times$     & $\times$     \\
Set up necessary software to run at boot & $\checkmark$ & $\checkmark$ & & $\times$     & $\times$     \\
Resize partition to fill whole SD card   & $\checkmark$ & $\times$     & & $\times$     & $\times$     \\
Install necessary software               & $\checkmark$ & $\checkmark$ & & $\times$     & $\times$     \\
Install software dependencies            & $\checkmark$ & $\checkmark$ & & $\times$     & $\times$     \\
Change hostname                          & $\checkmark$ & $\checkmark$ & & $\times$     & $\times$     \\
Change password                          & $\checkmark$ & $\checkmark$ & & $\times$     & $\times$     \\
Update one configuration file            & $\times$     & $\times$     & & $\checkmark$ & $\checkmark$ \\

\bottomrule
\end{tabular}
\end{table*}

\subsection{Secure Network Connectivity}\label{network-connectivity}

Every WiFi enabled device has to determine how to connect to the wireless network. This is not a problem for devices with keyboards and screens, but it becomes much more difficult when a device have none of these options. Currently, the state of the art solution for IoT devices is for that device to create a temporary wireless network. The person setting up the IoT device connects to the temporary wireless network, using the smart phone, selects the home network name and enters the password. This gives the IoT device the information it needs to connect to a WiFi network. This works with one or two devices, but becomes a nuisance when dealing with a few devices and increasingly untenable as the number of devices increases. To solve this problem, we create a novel protocol for our gateway to send the encrypted network name and password to wireless sensors using Ethernet source and destination address fields to encode the data. This greatly simplifies and speeds up the process of connecting sensors to a wireless network while integrating with existing WiFi protocols. The researcher can plug the sensor device into a power outlet and it automatically connects to the network and begins working with the deployed gateway.

\subsubsection{Out-of-Band Channel}

Sending the network name and password to an unassociated sensor is a challenge because data can not be sent directly to the sensor. The sensor is not associated with the access point, so it is unable to decrypt packets that are sent. However, the sensor is able to enter into monitor mode, allowing it to see the packets that are being sent, even if it cannot decrypt them. Instead of sending the data directly, data must be sent indirectly from an Ethernet connected device (gateway) to the unassociated device (sensors), through an out-of-band channel. One option would be to use timings to convey information. This is prone to errors especially since WiFi cannot guarantee when data will be sent due to CDMA. One part of the Ethernet frame that is not encrypted when it transitions to a WiFi frame is the address fields. We take advantage of this fact by encoding information in the address field. The idea is to have the gateway encode the network and password into the address fields of the Ethernet frames so that an unassociated sensor is able to decode the information.

Care must be taken by the gateway in how it encodes information into the Ethernet frame addresses. The gateway can change the source address because no one will be responding to the packet that is being sent -- the address does not have to be routable back to the gateway. However, the bytes in the MAC address have specific meaning that need to be followed or a frame risks getting rejected. The least significant bit of the first byte specifies whether the address is multicast (1) or unicast (0). The second least significant bit of the first byte specifies if the MAC address is globally unique (0) or locally administered (1)\cite{mac}. To follow this convention, we set the first two bits of the first byte to \texttt{10}. The destination address needs to be formed such that an access point will send out the packet on its wireless interface. There are a few addresses that guarantee this: broadcast (\texttt{FF:FF:FF:FF:FF:FF}), IPv4 multicast (\texttt{01:00:5E:xx:xx:xx})\cite{rfc1112}, and IPv6 multicast (\texttt{33:33:xx:xx:xx:xx}) \cite{rfc2464}. The IPv6 multicast address gives the most unused bytes, so we select this address.

\subsubsection{Header}

\begin{figure}[!t]
    \centering
    \includegraphics[width=\columnwidth]{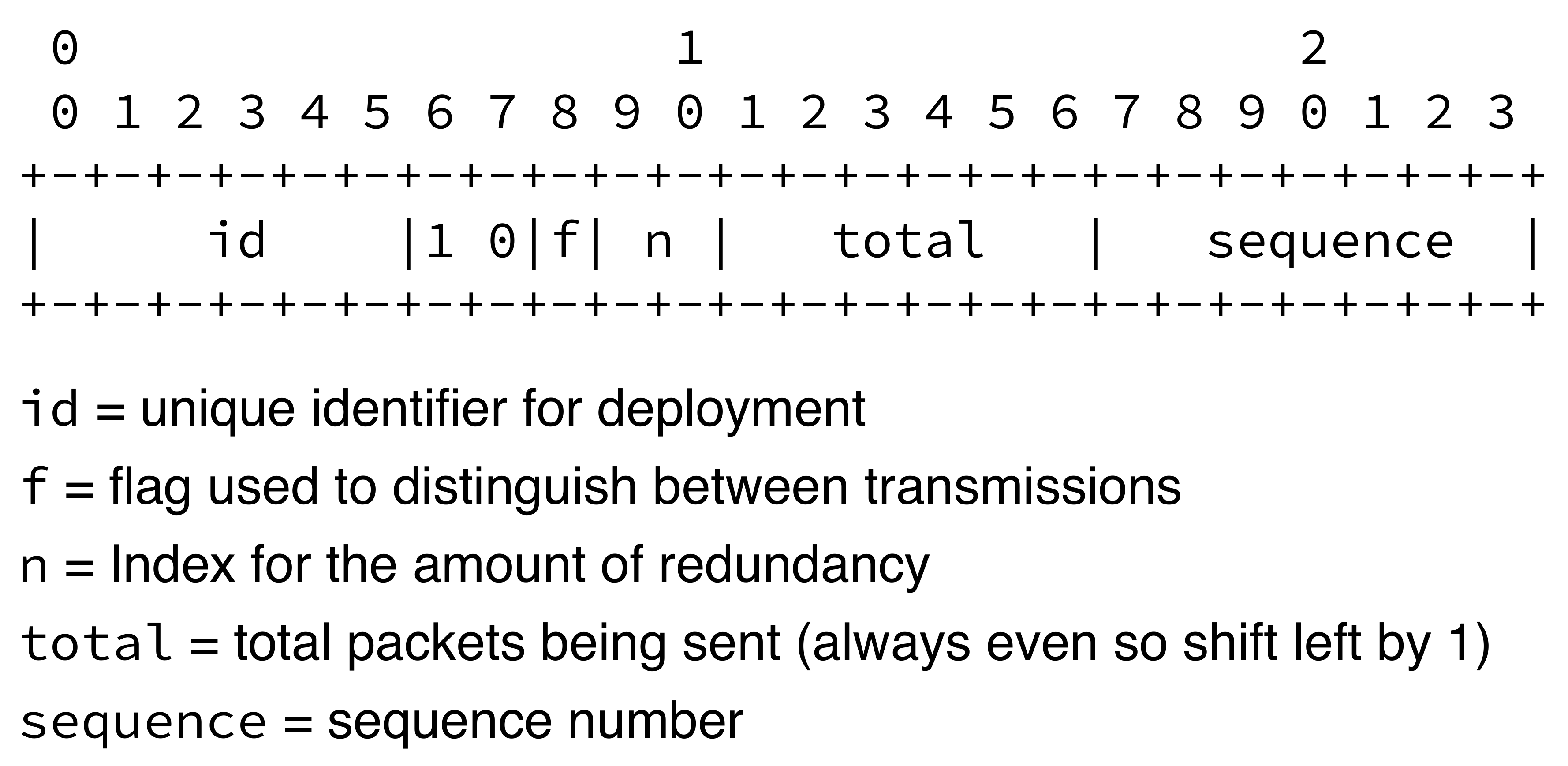}
    \caption{Outline of header format}
    \label{fig:header}
\end{figure}

Using the source and destination addresses as described above, there are almost 10 bytes that are available to encode data. We use the first 3 bytes of the source address as a header for our protocol and the rest of the bytes as the payload. Figure~\ref{fig:header} outlines the layout of the header. The first 6 bits are used as a unique ID to identify the gateway and sensors. Since each sensor will be in monitor mode, listening to all wireless communication on a certain channel, it is important that the sensor can filter out data that is unintended for it. Bits 6 and 7 set the MAC address to be locally administered and unicast, as mentioned above. Bit 8 is a flag used to distinguish between different transmissions. The flag is alternated between one and zero, allowing a sensor to know when one transmission has ended and another one has started. Bit 9 and 10 are used as an index into a predetermined array for how much erasure coding will be added. This is explained more in Section~\ref{fec}. Bit 11 through 16 is the total number of packets being sent. In the protocol, we make sure there is always an even number of packets being sent, so the total number of packets must be shifted left by one. The last 7 bits (bit 17 to 23) represent the sequence number of that packet. The rest of the 7 bytes are used to send the encrypted network name and password.

One thing to note is that only 126 packets can be transferred in one exchange due to the size of the total field in the header. The network name and password is small enough that this will not be a problem. However, if it were to become a problem, the information can be broken up into multiple exchanges and the data can still be transmitted.

\begin{figure}[!t]
    \centering
    \includegraphics[width=\columnwidth]{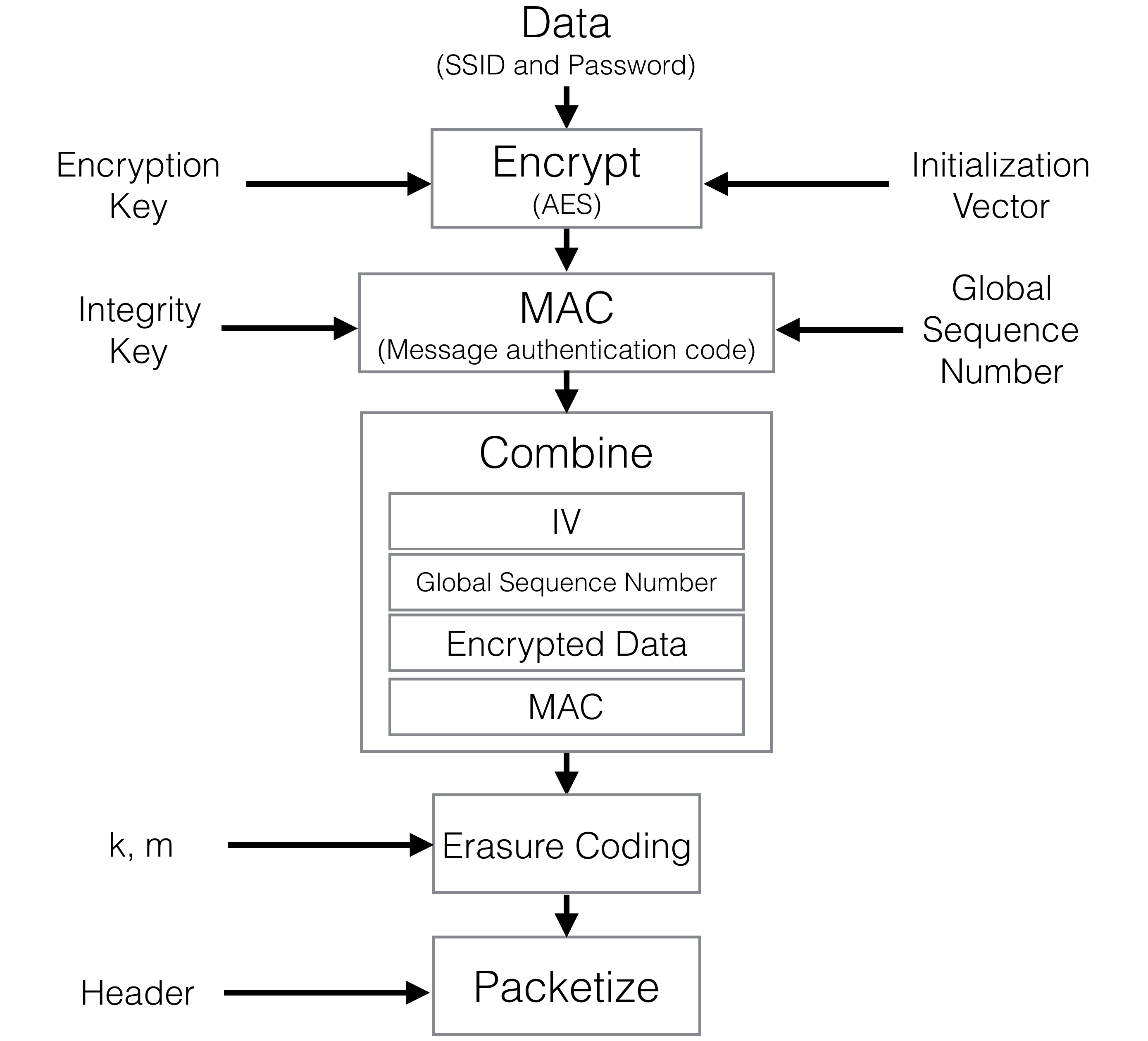}
    \caption{Flow of data through the protocol.}
    \label{fig:wifi-transfer}
\end{figure}

\subsubsection{Encryption and Integrity Protection}

To ensure that the network name and password are protected, we use AES-128 with CBC to encrypt the data. An initialization vector (IV) is generated and used to encrypt the data and is sent along with the encrypted payload. A ``global'' sequence number is added to the payload as well. We use the term ``global'' to distinguish between the sequence number that is part of the header described in Figure~\ref{fig:header}. We use the current time in the form of an epoch timestamp as our global sequence number since it is monotonic. A message authentication code is created based on the encrypted data and global sequence number. We use SHA256 as our hash for creating the message authentication code. This is added to the payload as well.

\subsubsection{Erasure Coding}\label{fec}

Erasure coding is the process of adding redundant data to the original data such that if there is data loss, the message can still be recovered. This is important for this protocol because we are using broadcast to send the packets, so there is no link layer acknowledgements. If there is some reasonable amount of loss at the receiver, we want the receiver to be able to decode the data.

We use Zfec~\cite{zfec}\cite{zfec-github} as our erasure coding algorithm. Zfec takes two parameters, $k$ and $m$. $m$ is the total amount of blocks that will be produced and $k$ is the number of blocks needed to construct the original message. For our purposes, we select $m$ such that the block size is equal to our payload size, 7 bytes.

In order to decode the data from Zfec, the receiver must know $k$ and $m$. $m$ is included in the packet header as the total number of packets. In order to save bits, we do not want to send $k$ directly. Instead, we use an array of predetermined values and send an index into that array. The values in this array represent the maximum loss tolerated while still being able to decode the message. For example, the array [.2, .3, .4, .5] would represent 20\% loss, 30\% loss, 40\% loss, and 50\% loss. If the gateway wants to support 40\% loss, it encodes the the data using Zfec such that $k = .4m$ and set bits 9 and 10 of the headers to 2. When a receiver receives a packet, it indexes into the array using bits 9 and 10, uses this value to calculate $k$ based on $m$, and decodes the message.

\subsubsection{Protocol}

To start the process of transferring the network name and password, the participant enters this information into a website that is running locally on the gateway. This information is encrypted, erasure coding is added, the data is packetized, and sent as an empty Ethernet frame with only the source and destination address set. Figure~\ref{fig:wifi-transfer} shows the general flow of data.

To ensure that all sensors receive the data, the gateway will continue to repeat this procedure, toggling the send flag, updating the global sequence number, and creating a new IV. The sensors enter into monitor mode and scan through all the channels, listening for packets from the gateway (using the ID segment of the header to filter out unwanted packets). Once a sensor has received enough packets to complete the message, it authenticates the message (using the MAC and global sequence number), decrypts the message, connects to the home access point, and notifies the gateway that it has connected. The website where the participant entered their information is updated live with the sensors that have connected. As time goes on, the gateway increases the amount of erasure coding it adds to the data. This helps to catch any sensors that have enough loss that they are unable to decode the data. Once all of the expected sensors have connected, the deployer can stop the gateway from sending the data.

\subsubsection{Adversary Model}

For this protocol, we assume that there is a trust relationship between the gateway and sensors and that two keys have been loaded onto the gateway and sensors before the deployment. This is done when setting up the software for the gateway and sensors. One of the keys is used for encryption and the other is used for integrity protection. We assume that an adversary can eavesdrop on the conversation and replay previously captured packets. The data is encrypted using a shared secret between the gateway and sensor so the adversary will be unable to decrypt the data. A new IV is used for each round of transmissions, so nothing about the encrypted data can be learned. The global sequence number protects against replay attacks, because the client can detect if the message is old. The adversary is unable to change the encrypted data or global sequence number without the sensor detecting it, due to the message authentication code.

\subsubsection{Alternatives}

Other methods exist to solve the problem of connecting a small device to a wireless network. The method of creating a temporary network is discussed above as an inconvenient option for large sensor deployments.

Another option would be to use another out-of-band channel for communicating this information. For example, a sensor could be equipped with Bluetooth just for the purpose of receiving this information. For our sensors, BBBs only have one USB port which is used for the WiFi adapter, so connecting up a Bluetooth adapter is not an option.

Another approach would be to use Wi-Fi Protected Setup~\cite{wps} which allows you to push a button on an access point, and on the device that you are connecting, to have them connect. Since we are using the participant's wireless network there is no guarantee that their access point would support this feature.

Lastly, and probably the simplest, would be to preload each sensor with the network name and password before deploying. However, this requires a participant entering their password in a survey taken before deployment. There are a few problems with this: 1) a participant may be uncomfortable entering in their password into a remote database; 2) a participant may not know their network name and password if they are away from their home; 3) there can be a high rate of errors when communicating the network name and password via survey. The cryptic passwords that come pre-installed on home access points might be hard to relay accurately over the phone or on a paper survey form (e.g., mistake a capital ``o'' for a zero). To make matters worse, a problem can only be detected after trying to deploy unsuccessfully. Without the instant feedback of knowing if the network name and password are correct, this system is not scalable.

\subsubsection{Evaluation}

\begin{table}[!t]
\renewcommand{\arraystretch}{1.2}

\centering
\caption{Packet loss on Beaglebone Black while using secure network connectivity protocol.}
\label{table:loss}
\begin{tabular}{@{}lrrr@{}} \toprule

         & \multicolumn{3}{c}{Percent Packet Loss} \\
           \cmidrule{2-4}
Location & Close & Medium &  Far            \\
1        &    37 &     70 &   93            \\
2        &    54 &     36 &   99            \\
3        &    90 &     99 &  100            \\

\bottomrule
\end{tabular}
\end{table}

\begin{table}[!t]
\renewcommand{\arraystretch}{1.2}

\centering
\caption{Packet loss on Beaglebone Black compared to Raspberry Pi using the secure network connectivity protocol at location 1.}
\label{table:pi-vs-bbb}
\begin{tabular}{@{}lrrr@{}} \toprule

                    & \multicolumn{3}{c}{Percent Packet Loss} \\
                      \cmidrule{2-4}
                    & Close & Medium &  Far            \\
Beaglebone Black    &    37 &     70 &   93            \\
Raspberry Pi        &   1.8 &   10.8 &  7.9            \\

\bottomrule
\end{tabular}
\end{table}

In developing our connectivity protocol, we collected data to understand the loss patterns in homes to inform us if erasure coding needs to be added. We ran the protocol, measuring the number of packets that were received by a sensor. We took these measurements from three distances from the access point: close (less than 2 feet away), medium (about 10 feet away), and far (about 20 feet away), using the BBB. We ran these experiments at three different deployment locations. Table~\ref{table:loss} shows the results.

We were surprised to find the amount of loss so high, even when close. To investigate this loss further, we ran the same tests at location 1, using the the same WiFi hardware and software, but instead of using a BBB, we use a Raspberry Pi. The loss is much lower as seen in Table~\ref{table:pi-vs-bbb}. This seems to show that there is problems with the drivers or the operating system on the BBB that make it unable to capture the packets. Even with the high packet loss, the BBB is the right choice for our research work and we were able to use erasure coding to compensate. We set the default tolerable loss to 60\% loss with the gateway able to adapt to 70\%, 80\%, or 90\% if needed.

\subsection{Sensor Discovery}\label{discovery}

To minimize the configuration at the sensor, they should not be required to know anything about the environment in which they are deployed. They are designed to passively collect data, waiting for a gateway to request data. This allows sensors to be deployment agnostic so that they don't need to be reconfigured for each home in which it is deployed.

Instead, the gateway discovers sensors that are on the network. To discover the sensors, we use CoAP's discovery multicast address and discovery URL. This allows the gateway to discover sensors on a predetermined IP address (224.0.1.187) and a predetermined URL (\texttt{.well-known/core}). In response, sensors send back their capabilities in the RFC 6690 link-format. This informs the gateway how the data will be formatted when sent by the sensor and the sensor's internal IP address. This IP address will be used by the gateway for all future communication. The gateway periodically sends discovery messages looking for new sensors that might have been added. The process of retrieving data from the sensor is described in Section~\ref{reliability}.

\subsection{Data Persistence}\label{persistence}

In our own implementations and through the experiences of others~\cite{hitchhiker}, it is known that power loss to devices is  inevitable  while sensors are operating in homes. Data loss can have a negative impact on a study and the trust of a system. Under these circumstances, data persistence is important to insure that no data is lost.

Using a database is a possible solution because it provides the data integrity that we need. However, using a database for an application like this is heavy and cumbersome. Also a database is not likely to be easy to implement in an embedded system environment.

A more lightweight approach would be to use some kind of queue that also persists the data. After investigating this option, we could not find a persistent queue that supports the work flow we were looking for. The required work flow is to ``peek'' at the data on the queue (copy the data without removing it from the persistent storage), upload the data, confirm that the data has been uploaded properly, and delete the data from the queue. All queue implementations that we found only supported pop/push semantics and did not support peeking or deleting (removing data from the queue without reading it). We feel like this work flow is universal enough that we can create our own persistent queue.

The sensor has two main processes: recording data from the sensors into the persistent queue and sending data to the gateway when requested. When a request for data comes in from the gateway, the sensor reads data from the persistent queue, sends it to the gateway, and deletes the data once receipt of data has been acknowledged. The gateway has a similar procedure for making sure all information that passes through it is saved to the database. By using a persistent queue, we ensure that all data collected  will be safe regardless of application failure or power loss.

\subsubsection{Evaluation}

\begin{figure}[!t]
    \centering
    \includegraphics[width=3.2in]{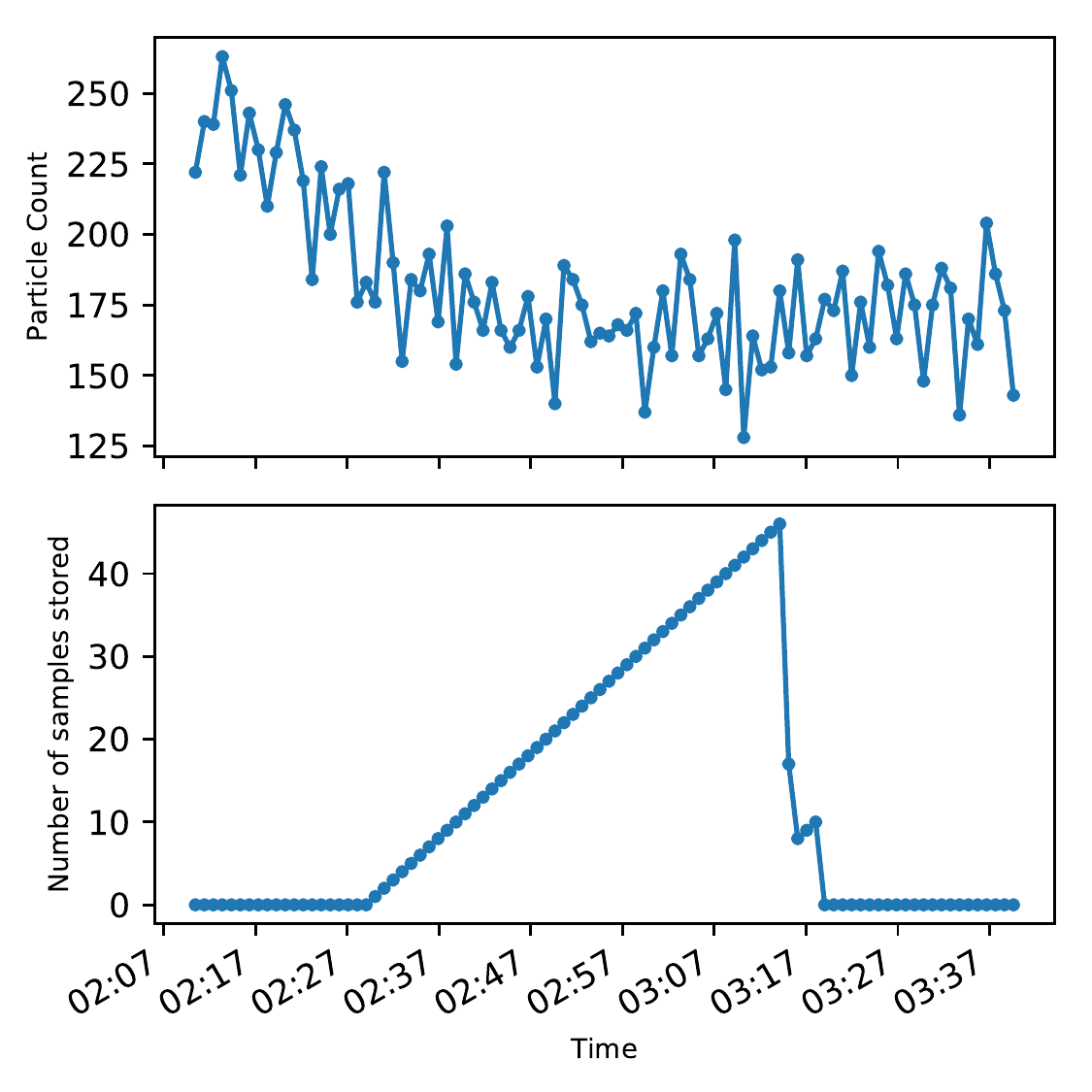}
    \caption{The top graph shows the particle count from a Dylos sensor. The bottom graph shows the number of data samples stored on the Dylos sensor.}
    \label{fig:gateway-loss}
\end{figure}

To evaluate the persistence mechanism of EpiFi, we set up a sensor connected to a gateway. We use a Utah Modified Dylos air quality sensor. A description of the Utah Modified Dylos air quality sensor is given in Section~\ref{modified-dylos}. We then disconnect the gateway from the network for about 40 minutes. This is to simulate an internal network error or a gateway getting disconnected or turned off by accident. Figure~\ref{fig:gateway-loss} shows the results. The top graph plots the particle count measurements that the Dylos sensor is taking. The bottom graph is the number of data points stored on the Dylos sensor. Both graphs have time as the x-axis. At about 02:28, the gateway gets disconnected and is unable to retrieve data from the Dylos. As a result, the amount of data points stored on the Dylos steadily increases. At about 03:15, the gateway gets reconnected and starts pulling data from the sensor. It continues to pull data until the number of samples goes back down to zero. This process takes about 4 minutes to complete. The samples from the Dylos sensor show no evidence that it was disconnected for 40 minutes.

\begin{figure}[!t]
    \centering
    \includegraphics[width=3.2in]{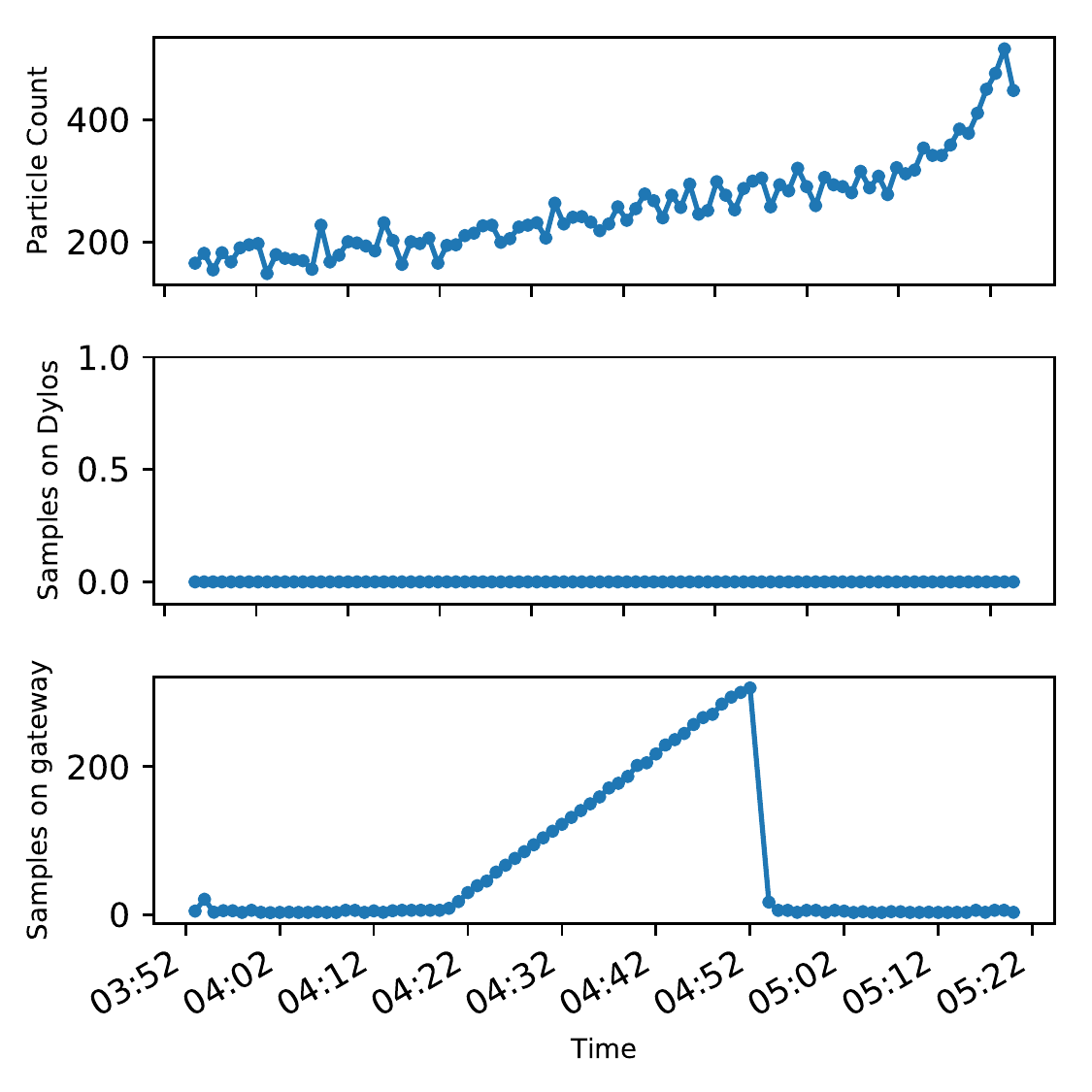}
    \caption{The top graph shows the particle count from a Dylos sensor. The bottom graph shows the number of data samples stored on the Dylos sensor.}
    \label{fig:internet-loss}
\end{figure}

Next we evaluate the data persistence and reliability when there is no Internet connection. Figure~\ref{fig:internet-loss} shows the results. The top graph is the particle count measurements that the Dylos sensor is taking. The middle graph is shows the number of data points stored on the Dylos sensor. Since the Dylos sensor and gateway are still connected, this value remains zero. The bottom graph is the number of data points stored on the gateway. All three graphs have time as the x-axis. Since the gateway is not connected to the Internet, the data is stored there until it is able to upload it. The gateway disconnects from the Internet at 4:21 and reconnects at 4:52. The process of uploading the backlogged data takes 3 seconds.

From both of these experiments, we can see that EpiFi is able to handle disconnections from sensor to gateway or from gateway to Internet. This is made possible through the data persistence component we created.

\subsection{Reliability}\label{reliability}

Even with data persistence, loss can occur when transferring between sensor and gateway and between gateway and database.  To ensure that data has been properly transferred, application level acknowledgements are needed. We designed a simple protocol on top of CoAP to be lightweight and flexible. Without such a protocol, there is no way for the sensor to know if the gateway received the data properly.

Our process for retrieving data from the sensors works in the following manner. After the gateway has discovered a set of sensors, it requests data from them in a round-robin approach. The gateway will make a GET request to a sensor for its data. Having the gateway request data in this manner removes the possibility of interference between sensors since only one sensor is ever sending data. The GET request has two important pieces of information: the number of data points the gateway wants from the sensor and the number of data points it is acknowledging from the previous request.

By having the gateway choose how many data points it wants from a sensors, the gateway is able to balance getting data quickly and scalability. The sensor sends at most the number of data points requested. If it does not have that many data points to give, then it gives the maximum it has. For example, if the request asks for 10 data points, but the sensor only has 5 data points stored, it will send the 5 data points it has.

\begin{figure}[!t]
    \centering
    \includegraphics[width=3.2in]{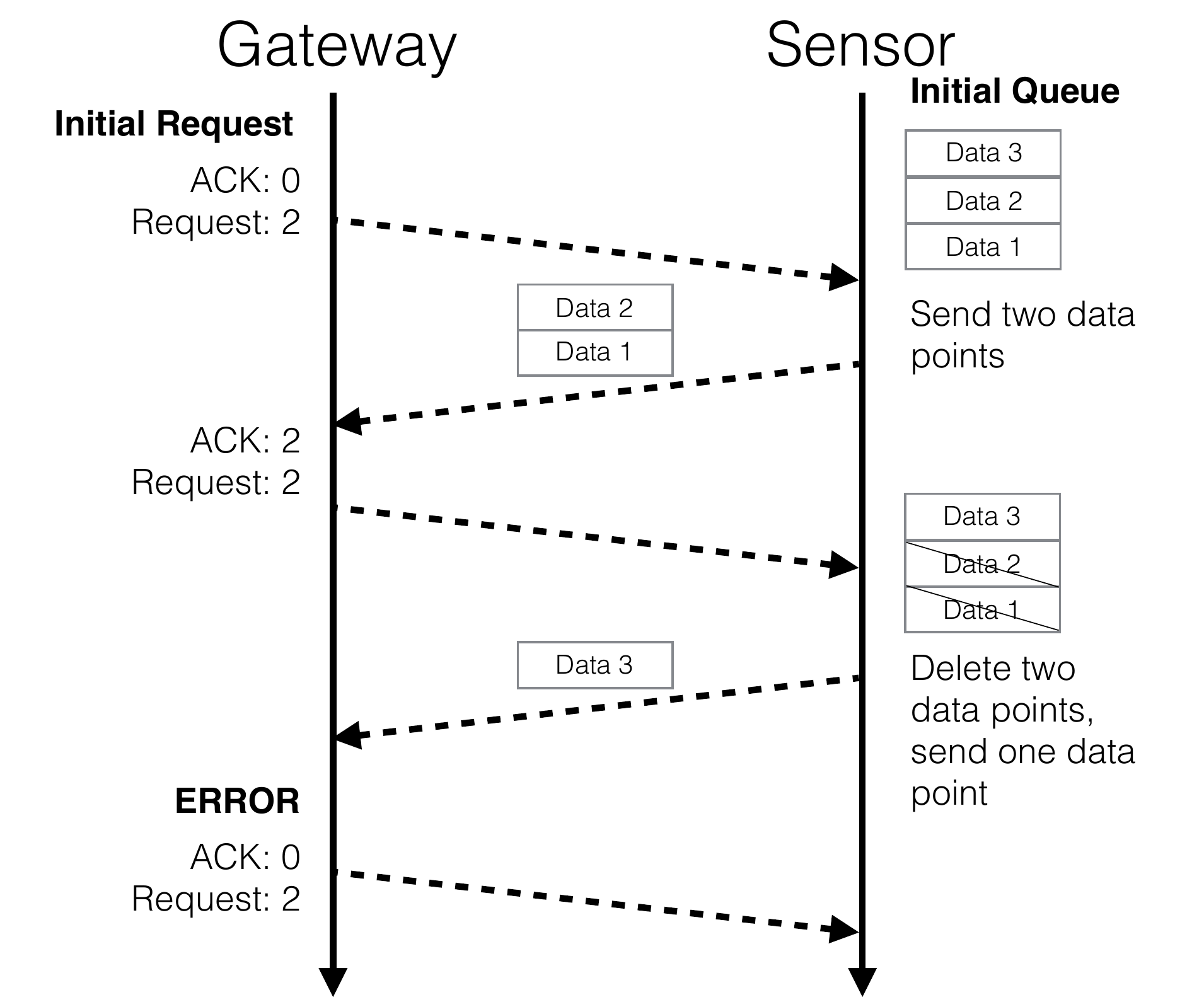}
    \caption{An example of packets exchanged between gateway and sensor.}
    \label{fig:gateway-sensor-exchange}
\end{figure}

The second portion of the request for data is the number of data points being acknowledged. The sensor has no way of knowing if the gateway has received \textit{and} processed the data properly. CoAP confirmable messages (ACKs) can be used, but this only tells us if the message was received properly and not processed. The number of data points acknowledged will typically be the number of data points received in the previous request, but if an error has occurred on the gateway, it sets this number to zero. In this case, duplicate data can be received. We are fine with receiving the same data twice rather than losing data inadvertently. Duplicated data is detected by the database, InfluxDB, and ignored~\cite{influxdb-dups}. When the sensor receives the number of data points acknowledged, it is free to delete those data points. Figure~\ref{fig:gateway-sensor-exchange} illustrates an example of this protocol.

By default, EpiFi will request 10 samples from a sensor at a time. This is a configuration option, that a researcher can change, which depends on how often the sensor records measurements and how big the payload is. To ensure fairness between sensors, the gateway will request at most 120 samples from a sensor before moving on to the next sensor.

The reasoning behind this is to protect sensors from being temporarily starved. If one sensor has been disconnected for a long time, it will have a large backlog of data. Without this mechanism the gateway will continue to pull data from it until all of the data is gone, ignoring the other sensors while it does so. Rather than emptying out the backlog before moving on to other sensors, it will stop requesting data after it has received 120 measurements and move on to other sensors. This provides more even queue lengths and latency across sensors. This is configurable by the researcher and depends on the deployment needs.

\subsubsection{Evaluation}

See Section~\ref{large-deployment} for evaluation on reliability.

%% file: parts/deployments.tex
\section{Deployments}\label{deployment}

To verify our architecture, we deploy our system in four different studies. With the exception of one, each study involves deployment in multiple homes. We worked with pediatric asthma researchers who study the relationship of exposure to indoor air pollution and the symptoms and treatment of children with asthma. We also worked with air pollution scientists who understand the chemistry and appropriate sensors for indoor air quality measurements. Together, with these domain experts we design and deploy the experiments. For these studies, we collect data from sensors, primarily air quality sensors, that we integrate into EpiFi.  We first describe the sensors we used and then describe each deployment. These deployments illustrate the abilities of EpiFi. We also discuss lessons learned from the deployments.

\subsection{Sensors}

To measure the air quality in homes, we used two air quality sensors: the \emph{Utah Modified Dylos Sensor} and the \emph{AirU}. We also use a variety of other commercial off-the-shelf sensors of various modalities, wireless interfaces, and protocols to show the capabilities of EpiFi to integrate with consumer product sensors.

\subsubsection{Utah Modified Dylos}\label{modified-dylos}

The Dylos DC1100 air quality sensor is a commercially available sensor that measures the concentration of airborne particulate matter \cite{dylos}. It pulls air through the sensor chamber with a fan, and uses a laser and image sensor to estimate when a particle passes through the chamber and its approximate size.  The DC1100 sensor provides two measurements of particle count, small and large.  The small particle count is a count of particles larger than 0.5 $\mu$m, while the large particle count is a count of particles larger than 2.5 $\mu$m.  In air quality science, the standard measure, $PM_{2.5}$, is a measure of the mass concentration per unit volume of all particles smaller than 2.5 $\mu$m.  The difference between the large and small particle count is a measure that is considered to be approximately related to the $PM_{2.5}$, although the relationship varies for different particles because the mass of the individual types of particles.  The Dylos has been used in multiple research studies because of its low cost (USD 200), which is 10-100 times lower than laboratory-quality measurement systems. Although many commercial air quality sensors exist, the calibration and characteristics of the Dylos have been extensively studied and reported on \cite{semple2012inexpensive}.

To integrate the Dylos into EpiFi, we connect the serial port of the Dylos to the BBB. Next, we connect a temperature and humidity sensor to the BBB using the I$^2$C protocol. Last, we replace the LCD screen of the original Dylos with a RGB LCD screen and connect it to the BBB. The LCD screen is used to display air quality readings and other diagnostic information.

We integrated three software sensors into the Dylos. These include a sensor to measure the wireless characteristics of the Dylos, a sensor that pings the gateway every five seconds to measure the latency and any errors, and a sensor that pings the a remote server every 15 seconds to measure the latency and any errors. We use these sensors for debug and for recording system performance.

\begin{figure}[!t]
    \centering
    \includegraphics[width=\columnwidth]{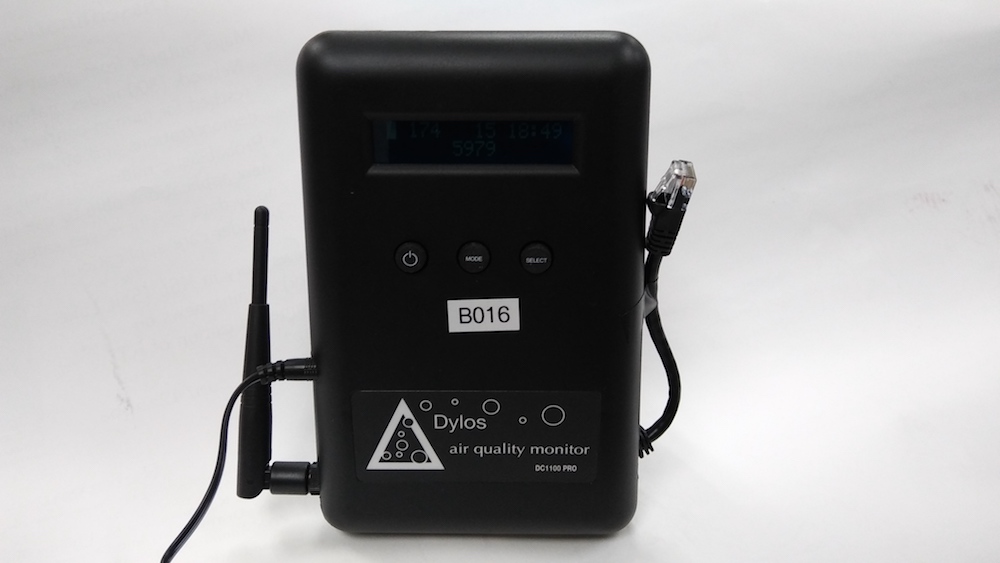}
    \caption{Utah Modified Dylos air quality sensor used in deployments.}
    \label{fig:dylos}
\end{figure}

The Dylos updates the air quality reading on the serial port every minute, so we set up EpiFi to pull data from these sensors once per minute. We modified the case of the Dylos so that we could fit the BBB and temperature sensor inside with space for a WiFi antenna and Ethernet cable to stick out. Figure~\ref{fig:dylos} shows the Utah Modified Dylos.

\subsubsection{AirU}

\begin{figure}[!t]
    \centering
    \includegraphics[width=\columnwidth]{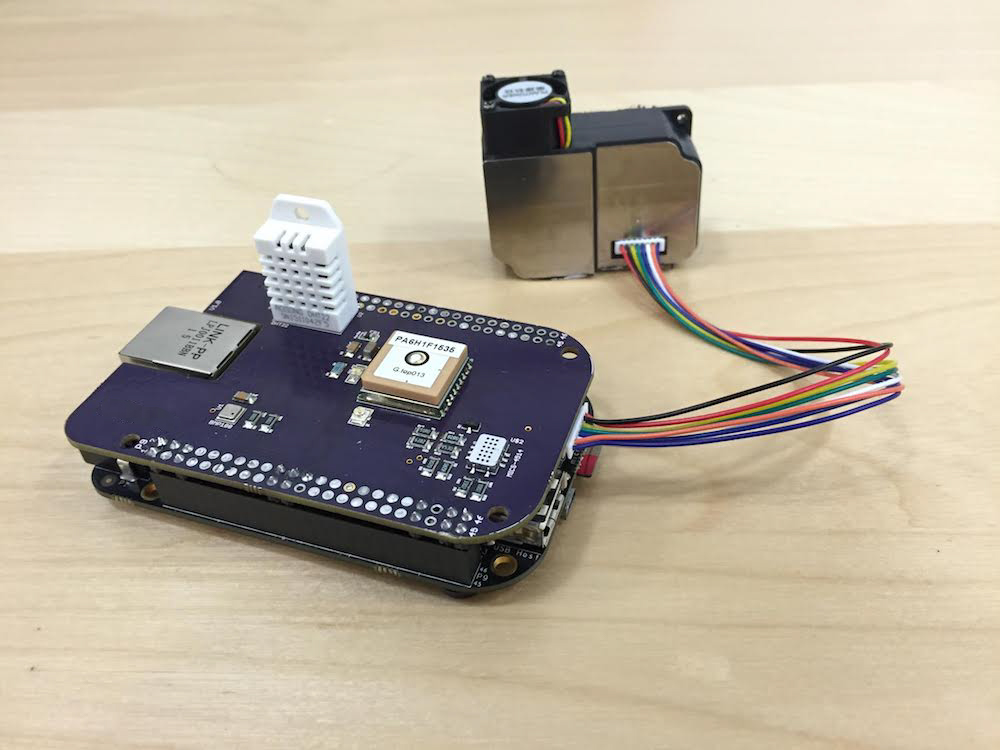}
    \caption{AirU air quality sensor used in deployments.}
    \label{fig:airu}
\end{figure}

The AirU~\cite{airu} is a beaglebone cape which contains an air quality sensor (Plantower PMS3003), temperature/humidity sensor, and GPS sensor. For our deployments, we do not use the GPS sensor because the sensors are indoors and the GPS readings are unreliable. The Plantower air quality sensor reports three values every 60 seconds. These values correspond to PM$_1$, PM$_{2.5}$, and PM$_{10}$. Figure~\ref{fig:airu} shows an example of this device.

\subsubsection{Other Sensors}

For one of our deployments, we use an Aeotec MultiSensor and WeMo motion sensor to detect movement in a home and a Schlage door sensor to detect when a door is opened and closed.

\subsection{Large Deployment} \label{large-deployment}

For the first type of deployment, the experimenters wanted to study how air quality differs across space and time inside of a house. Most studies involving indoor air quality deploy one sensors in the home. The goal of this experiment, for the domain experts, is to learn in more detail how the air quality is a function of room, what caused the air pollution, and where that pollution originated.  An important question, for future studies, was about determining the benefit of having multiple sensors vs.\ a single sensor, and generally, knowing how many sensors and at what locations are required to get an accurate picture of a home's air quality. For this deployment, we set up eight Dylos sensors in various rooms in the house and one sensor outside. Beyond measuring day-to-day living, controlled experiments were performed, such as opening up the front door for a certain amount of time, or lighting a candle. We deployed this system in two homes.

\begin{figure}[!t]
    \centering
    \includegraphics[width=0.95\columnwidth]{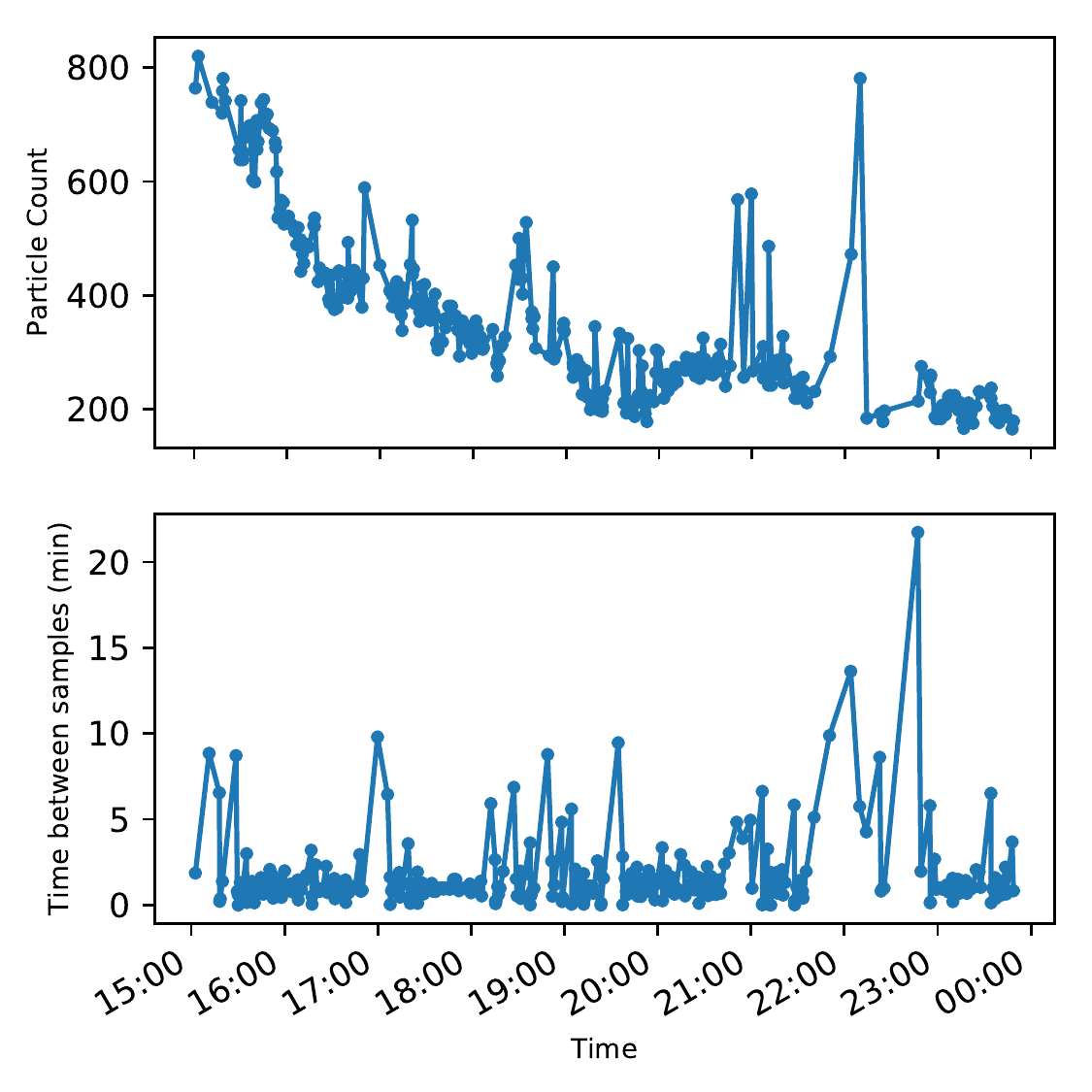}
    \caption{The top graph shows the particle count from a air quality sensor. The bottom graph is the time difference between samples. The time difference should be constant.}
    \label{fig:missing-data}
\end{figure}

This was the first deployment of EpiFi. From this deployment, we learned many valuable lessons on how to improve the architecture. We originally had the gateway set up as an access point that all sensors connected to. We found that sensors deployed in rooms far from the gateway could not connect to the gateway due to a weak signal strength. We also saw packet loss much higher than we were expecting. Figure~\ref{fig:missing-data} shows an example of such a data collection. The top graph shows the particle count as recorded by one of the sensors compared to time. Data samples should be every minute. The bottom graph shows the time difference between the gateway's received measurements. Both graphs have time as their x-axis. You can see that there is a great deal of variation, and there are times when no data is received for more than 20 minutes.

From this experience we learned two things. First, by switching to the home's wireless access point, the amount of loss is significantly reduced. Second, we learned the importance of distinguishing between packet loss and data loss. Packet loss can occur because of the nature of the wireless medium that we are using. Data loss is when a measurement collected by a sensor is lost completely. To eliminate data loss, we implemented persistence (Section~\ref{persistence}) and reliability (Section~\ref{reliability}). Before making these changes, we were seeing an average of 25.57\% data loss for each sensor and after making these changes, we are seeing 0.0\% data loss.

\begin{figure}[!t]
    \centering
    \includegraphics[width=0.95\columnwidth]{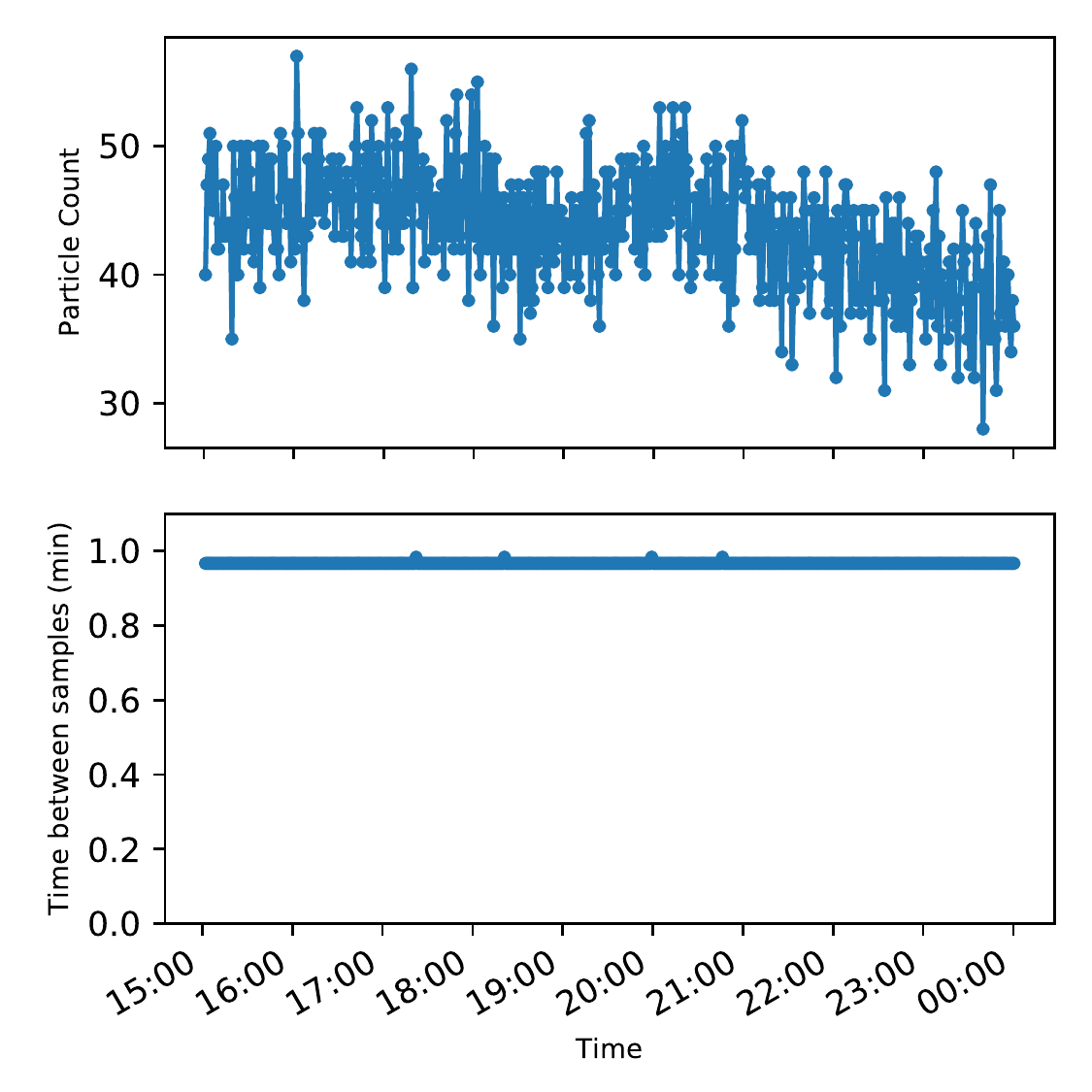}
    \caption{The top graph shows the particle count from the air quality sensor. The bottom graph is the time difference between samples. The time difference is constant.}
    \label{fig:good-data}
\end{figure}

A more subtle problem we discovered in this deployment was the importance of recording a timestamp when the measurement was taken and sending it with the data. Even without any data loss in the network we found that the measurements we were taking were not evenly spaced out as we expected. We originally were sending the data without a timestamp and when the data got to the gateway, it would timestamp the data and upload it to the database. This had the advantage of only dealing with one clock, the gateway, which we could ensure was accurate. There is enough latency and jitter in the wireless network that it had an effect on the data. We also found that two measurements that were taken at the same time, such as the Dylos sensor returning a large and small air quality reading, would have slightly different timestamps. This was due to the way Home Assistant processes incoming data. From a research point of view, this made it difficult to compare corresponding measurements, because two measurements that happened physically at the same time, would have slightly different timestamps. We fixed these problems by recording the timestamps when the measurements are being made and sending this timestamp with the measurement. Our custom component for uploading data then uses this timestamp for all measurements that were included with it. Figure~\ref{fig:good-data} shows an example of data where these problems have been fixed.

During the two month period that this deployment was in one of the homes, each Dylos sensors lost power on average 3 times. In spite of the power losses, no data was lost and a deployer did not need to come back to the home to set up the sensors again. Once the sensors were powered back on, normal operation continued.

% Query to get power loss data:
% SELECT COUNT(value) FROM sequence WHERE home_id = 'prisms_house_4' AND value = 1 AND time >= '2017-01-21T04:40:21.094514944Z' GROUP BY entity_id

\subsection{Furnace Deployment}

Our second type of deployment uses air quality readings to control a home thermostat fan. Turning on the fan to a home furnace pulls air through the furnace filter, which helps to remove pollutants, improving the air quality.  The goal of this deployment type is to explore the trade off between using the fan to improve air quality, and the energy cost of increased use of the furnace fan. For this deployment type, we use two air quality sensors: one in a bedroom and one in the main living area. These two locations typically are places where people spend the majority of time in their home.

We built a custom component for Home Assistant that reads data from the air quality sensors and if the air quality goes above a certain threshold, it turns the fan on. After a certain amount of time, the fan is turned off. The custom component also tracks the thermostats regular schedule and input from the user as not to disrupt normal functioning of the fan. This system has been deployed in two homes.

A lesson we learned from this deployment is that different applications have different requirements for real-time data. For this deployment, we want the data to be as recent as possible. However with other types of deployments, data that is minutes old might be fine. As a result, we made the speed at which the gateway pulls data from the sensors a configuration option. The original default (used by the large deployment) was five minutes. For this deployment, we set the gateway to pull data every 15 seconds. This ensures that we will get data at the most 15 seconds after it is measured.

\subsection{Clinical Deployment}

The third type of deployment is designed by pediatric asthma researchers.  There is significant evidence that outdoor air pollution levels can be used to predict asthma exacerbation \cite{pope2000epidemiology}.  Further, incorporating outdoor levels into treatment, including decisions about when to talk with a nurse and the appropriate dosage of medication, can reduce the frequency of emergency department visits \cite{asthma}.  However, such studies do not have accurate measures of an individual asthma patient's exposure to pollution.  First, these studies use only outdoor air measurements, and further, they use the measurements from outdoor monitoring stations, typically one sensor per city, which may be located far from the air to which an individual is exposed.  The type of deployment study reported here is motivated and designed by pediatric asthma researchers interested in sensor data collected closer to the patient and its ability to improve treatment of asthma.

For this study, researchers identify three locations for air quality sensors:  sensors in the bedroom and main living area to measure the air quality where people send the majority of their time when they are at home, and an outdoor air quality monitor to measure the air quality immediately outside of the home, to be compared with data from the city-wide sensor.

% Show data or don't show data see how behavior changes
As part of this study, the researchers want to understand how being presented with home air quality in your house affects behavior. They split up the study into three phases. In phase 1, the participant has no feedback about the air quality in the house. In phase 2, the participant has access to the air quality readings from a web interface. The user must check the web interface. In phase 3, the participant receives a text notification with a link to a survey when the particulate matter level in the air goes above a certain threshold. This survey asks questions about what activity they were doing and shows a graph of the data for the past hour. EpiFi is designed to enable each of these three phases. Further, study participants answer daily and weekly survey questions designed to assess asthma symptoms.

As part of these deployments, epidemiologists involved with the study deployed our system in homes. This gave us valuable insights into how we can make deployments easier and what was not working with early versions of EpiFi. After talking with these deployers, we learned the importance of having some kind of feedback from the sensors. This feedback is crucial for deployers who find they need to debug the sensors. For example, if there is some kind of sensor hardware failure, it is important for the deployer to know this as they are deploying rather than after the fact, when they are looking at the data. To solve this problem, we extended our Sensor class interface to support the \texttt{write} method which allows sensors to receive data as well. The LCD screen on the Utah Modified Dylos is an example of such a sensor.

\subsection{Exposure Deployment}

The last deployment shows the ability of EpiFi to support a variety of sensors and integrate them together. The motivation for this deployment is to improve the ability of a system to quantify a person's exposure to bad air quality in their home.

We use motion sensors next to the air quality sensors to determine if a person is close to the sensor or not. This allows us to have a better idea of what kind of exposure a person is experiencing. For example, someone might burn dinner, causing the air quality to drop in the main living area. The person then might go into their bedroom. Without knowing the true location of the person, you have to come up with an estimate of their exposure based on the air quality readings of the sensors (i.e. take an average of the sensors or take the maximum value). Putting a motion sensor next to the sensor gives us a better idea of the participant's true location so we are able to make better estimates of their exposure.

We deployed two air quality sensors in a home. We use a WeMo motion detector that uses WiFi and an Aeotec MultiSensor that uses Z-Wave. We use a Schlage Z-Wave door sensor to know if the door is open or closed. Using the motion sensor and door sensor, we can infer when someone is home or not. All of this data gets combined by EpiFi and uploaded to the database for analysis. Using this data, researchers may be able to obtain a more accurate measure of a person's exposure to air pollutants.

%% file: parts/conclusion.tex
\section{Conclusions and Future Work}\label{conclusions}

In this paper, we described EpiFi, an architecture for epidemiologists to use to build and deploy experiments. The design of EpiFi is targeted to address issues that are unique to epidemiologists. The components address real and important problems when dealing with deployments in study participants' homes, such as configuration, secure network connectivity, sensor discovery, data persistence, and reliability. These components work together to create a flexible and robust system, allowing epidemiologists to focus on epidemiology rather than networking. Being able to easily deploy sensors in homes will allow researchers to gain insights about the relationships between exposure and chronic diseases.

In the future, we would like to explore additional wireless protocols as alternatives to WiFi. One promising wireless technology is LoRa~\cite{lora} a wide-area wireless network designed specifically for IoT devices. In EpiFi, the gateway could be a LoRa base station with enough range to easily cover a house, an apartment building, or even a commercial building. This would also help EpiFi to be able to be deployed in more locations where WiFi is not an option.

EpiFi can benefit from more management tools. In the future, we would like to build a web interface for researchers to monitor their devices and add the ability to notify researchers when a sensor has gone down or something else has gone wrong with a deployment.

Lastly, an increase in the hardware support for sensors would be beneficial. Although the gateway allows for many different types of devices to interact with it, currently EpiFi only supports a Beaglebone Black as the foundation for a sensor. To enable a new class of sensors for EpiFi, additional support for WiFi microcontrollers such as the NodeMCU ESP8266~\cite{esp8266} would be beneficial.